\documentclass[aps,pre,twocolumn,superscriptaddress]{revtex4}
\usepackage{tabularx,graphicx}
\begin{document}

\title{Adaptive learning and coloniality in birds}
\author{M.~A.~R.~de~Cara}
\email{angeles.decara@icmm.csic.es} 
\affiliation{Instituto de
Ciencia de Materiales (CSIC). Cantoblanco, E-28049 Madrid, Spain.}
\author{O.~Pla}
\email{Oscar_Pla@amsinc.com} 
\affiliation{Instituto de Ciencia de
Materiales (CSIC). Cantoblanco, E-28049 Madrid, Spain.}
\author{F.~Guinea}
\email{paco.guinea@icmm.csic.es} 
\affiliation{Instituto de Ciencia
de Materiales (CSIC). Cantoblanco, E-28049 Madrid, Spain.}
\author{J.~L.~Tella}
\email{tella@ebd.csic.es}
\affiliation{Estaci\'on Biol\'ogica de Do\~nana (CSIC).
Avda. M. Luisa s/n,
E-41013 Sevilla, Spain}

\begin{abstract}
We introduce here three complementary models to analyze the role
of predation pressure in avian coloniality. Different explanations
have been proposed for the existence of colonial breeding behavior
in birds, but field studies offer no conclusive results. We first
propose a learning model in which the decision of birds are taken
according to the collective performance. The properties of the
system are then studied according to a model in which birds choose
according to their individual experience, and the agreement of the
introduction of spatial structure with field data are then shown.
\end{abstract}
\pacs{
87.23.-n 
05.65.+b 
89.65.-s 
89.75.-k 
}

\maketitle

\section{Introduction\label{s:intro}}


In the last few years there has been an increasing interest
in the understanding
of learning processes of collective behavior,
specially in systems of interacting agents.
The aim of these studies is to reproduce qualitatively features of economic
or biological systems~\cite{Aetal88,S96}.

Colonial breeding behavior in birds has been extensively
studied~\cite{L68,SK90,Retal98}.
During the breeding season
vertebrate social systems can be behaviorally classified into
three main groups: territorial, cooperative and colonial. This
classification is according to the genetical relationship with the
other members of the colony, and to the spatial organization.
While territorial and cooperative behavior are evolutionary
understood, colonial behavior remains an open question.
Different hypotheses have been
put forward in order to explain this behavior, like minimizing the
distance required for foraging~\cite{H68}, observation of conspecific
foraging groups~\cite{T56,BB96,BB01},
information transfer at the colony~\cite{WZ73},
shortage of nests~\cite{S76,SC87}, or predation pressure~\cite{L68}.

One of the difficulties in verifying the previous hypotheses is
that present day conditions need not to coincide with those which
lead to colonial behavior in the first place. Thus, modelling of
bird populations using reasonable assumptions for bird behavior
can be useful in the elucidation of possible scenarios favorable
towards the evolution of coloniality.

Some theoretical studies give
support to the hypothesis that information transfer at the
colony increases the tendency towards
colony formation~\cite{BL88,BS95,B97}.
However this hypothesis requires the previous existence of a group or colony,
and therefore it cannot explain by itself the evolution towards colonial behavior.

Predation can induce colonial habits in many ways. The simplest
passive mechanism is the dilution effect provided by a colony
of sufficiently large size~\cite{V77,B81,T96}.
In addition, the detection
and defense capabilities are enhanced
in colonies~\cite{K64,HS76,V77,A95,T96}.
On the other hand, the lack of significant predation pressure on
some colonial species has been used as evidence against the
predation hypothesis~\cite{B81,F89}, although a phylogenetic analysis
of coloniality across bird species shows a strong correlation with
exposure to predation in the past~\cite{Retal98}.

The present work analyzes the role of predation on the formation
of colonial habits by means of a mathematical model which incorporates
some of the known facts about the response of birds to
attacks by predators, and makes simple assumptions about the memory
and learning processes at play.

\section{Modelling learning processes in social systems}


\subsection{Minority Game and Individual Minority Game}

We model birds experience in a similar way to that used in the
``Minority Game'' (MG) model~\cite{CZ97,Z98,Jetal98,Cetal99,web}.
The minority game was introduced in the analysis of decision
making by agents with bounded rationality, derived from ``El
Farol'' bar problem~\cite{A94,CZ97}. The model describes $N$
agents which must make a choice between two alternatives (originally
defined as $\{0,1\}$, later on described as $\{-1,1\}$). Agents
make choices using as input the preceding collective performance. A
successful choice for an agent is that which
no more than half the total of agents choose.
The bounded capacity of each agent is modelled assuming that
agents can only remember the last $m$ rounds of the game. This
time span defines $2^m$ possible outcomes which each agent needs
to consider. Hence, the number of strategies that the agents can
use is $2^{2^m}$,
each of which is a set of choices for each possible
previous outcome.
To keep the assumption of bounded rationality,
agents have a limited number
$s$ of strategies, taken at random (in most
studies, $s=2$). Each strategy has an independent score, which is
updated after each move, according to its performance.

The game is defined by three parameters: $N$, the number of
agents, $m$, the number of time steps agents use to determine the
next decision, and $s$, the number of strategies available to each
agent. Depending on the ratio $\alpha=2^m/N$, a phase transition
has been found~\cite{Savit99,CMZ,GMS00,HC01}. This transition has
been successfully analyzed using replica symmetry breaking
and dynamical mean field theory
methods~\cite{CMZ,HC01}. The phase at $\alpha \ll 1$ has many agents
which are able to identify the ``optimal'' strategies, leading
to a poor global performance, as the minority group is usually
small. For $\alpha \gg 1$, agents play almost at random, as the
total number of strategies being used is a small fraction of the
total number possible. Near the phase transition, there is a
situation where groups of agents tend to play anticorrelated
strategies, and the global performance has a maximum.

The model has been generalized in different directions (see~\cite{web}).
There are different versions in which the agents are allowed to
evolve.
The strategies with which each agent is endowed can be
considered its ``genotype'', and can be allowed to change.
Alternatively, each agent can have an extra character, which allows
it to favor a given strategy or its opposite~\cite{LHJ00}.

The main objective of the present work, as described previously,
is the modelling of processes which lead to individual or
collective behavior and which are determined by responses to
unexpected events, like predation. While the generic pattern of
response can be genetically defined, we will concentrate on the
adaptation to the habitat which takes place during the lifetime of
each individual, using the agent's past experience only. Hence,
our starting point will be a variation of the minority game in
which the information used by each agent is not taken from the
collective history, but from the agent's own choices and
perfomance~\cite{Cetal00}. We define this version of the minority
game as the individual minority game. Previous studies show that
the use of different sources of information by different agents
change significantly their behavior in the $\alpha \ll 1$ region,
where some of the ``herding'' effects described in this section
are avoided. On the other hand, the fact that the information used
by each agent cannot be considered a random input~\cite{C99},
makes it difficult to use analytical techniques.
\begin{table}
\begin{ruledtabular}
\begin{tabularx}
{0.4\columnwidth}{%
>{\centering\arraybackslash}X
>{\centering\arraybackslash}X}
\mbox{Signal}&\mbox{Decision}\\
\hline
11&1\\
\hline
10&1\\
\hline
01&0\\
\hline
00&1\\
\end{tabularx}
\end{ruledtabular}
\caption{Example of strategy, for $m=2$\label{strategy}}
\end{table}

\subsection{Coloniality by predation}

In our approach to colonial behavior
we assume that the available choices to birds are limited to two possibilities
each breeding season:
they can either form an individual nest, or join an existing colony.
Therefore we find a binary system of decisions,
equivalent to the MG model.

Successful breeding individuals
tend to be faithful to their previous nesting site.
Birds choose a colony or an isolated nest depending on their previous
experience.

Each season, birds can be predated with probability $p$. We take as
our unit a breeding couple. A predation event does not imply
the birds which form the predated pair are removed from the
population,
but that the nest suffers predation from small
animals (rats, snakes, etcetera) which eat or damage the eggs.
Therefore, the reproductive success of the couple is zero or small,
and they learn from the experience.
These ``small'' predation events are much more common in nature
than those which involve big predators, such as mammals, which can destroy
the whole colony.
The birds have a finite
lifetime, which limits their ability for learning (see below),
which is not related to predation. The objective of
the present work is the study of simplified learning schemes
by which birds can aggregate in colonies,
and this learning mechanism can only occur when birds survive
to predation.
Hence,
``big'' predation events are irrelevant for this purpose.

Each bird has, as already mentioned,
$s$ strategies (see table~\ref{strategy} for example of a strategy
of memory $m=2$).
Each of these strategies has a score, which reflects
the innate preferences of each bird, or
the degree of reproductive success that the bird would have had if it had
followed it.

We consider different scoring mechanisms, as the available
biological data can be interpreted in different ways. We first
assume that this score is updated using collective information
from the performance of all individuals. Next, we analyze the case
when each bird updates the scores of its strategies using
information from its own success in previous occasions. Hence,
each bird uses information different from that used by other
birds, and the model departs from the minority game usually
analyzed in the literature, where the score of strategies is the
same for all agents.

The finite lifespan of the birds is modelled with a probability of
setting the scores of the strategies of a bird to zero
$p=\frac{1}{v}$ at each time step, where $v$ is the average
lifespan of the birds. This is equivalent to introduce a new bird
with no previous experience, and allows us to keep the population
size constant.

\section{Results~\label{s:model}}

\subsection{Collective scoring model\label{ss:mf}}
We first study a scoring scheme in which the
collective traits used in the standard minority game are combined
with the use of information private to each individual. The scores
are updated according to the following rules: i) The scores
corresponding to strategies which lead to the outcome not chosen
by the bird are updated according to their average success among
the birds which have followed them, ii) The scores corresponding
to the strategies which lead to the actual choice taken by the
bird are updated according to the success obtained by the bird at
that season. We assume that, in the absence of predation, the
innate tendency of the birds is such that the score assigned to
strategies leading to individual behavior is twice that for
colonial behavior.

This choice of scores takes into account the innate tendency of
many birds towards an individual behavior~\cite{T96}. The dilution effect, which
favors colonial behavior in the presence of predation, is included
in two ways: i) birds which choose an individual behavior
update the scores of the strategies leading to colonial
behavior taking into account the dilution effect which
exists in a large colony which includes all the birds and ii)
birds which choose a colonial behavior are predated less
often, depending on the size of the colony to which they
are in, which we assume to include all colonial birds.

\begin{figure}
\includegraphics[width=\columnwidth]{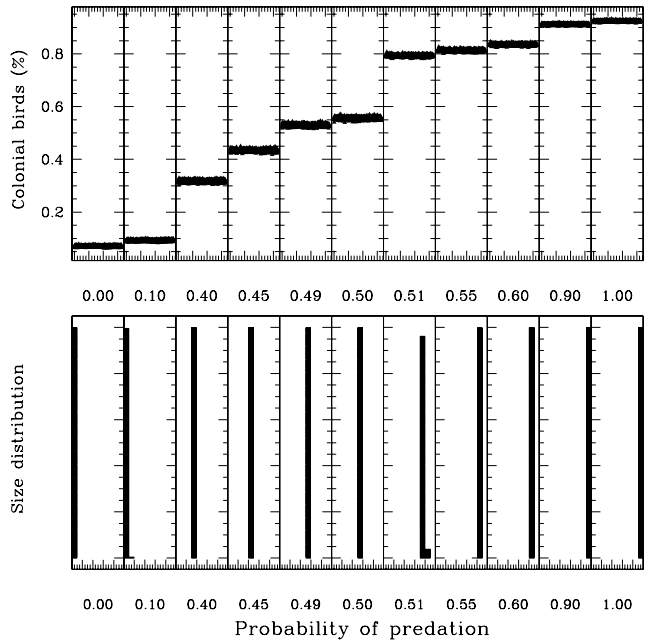}
\caption{Results for $m=2$, $s=5$, $N=1000$ and $v = 9$.
Top, temporal evolution of the model for
different predation pressures, from $t=1$ to $t=16384$,
Bottom, distribution
of colony sizes for the same
temporal evolution.}
\label{f:mf1}
\end{figure}

We have studied this model for different values of the parameters,
and we have found that the same qualitative features as
we vary the number of strategies $s$ available to each bird or
its memory $m$. Typical results are shown in fig.~\ref{f:mf1}.

\begin{figure}
\includegraphics[width=\columnwidth]{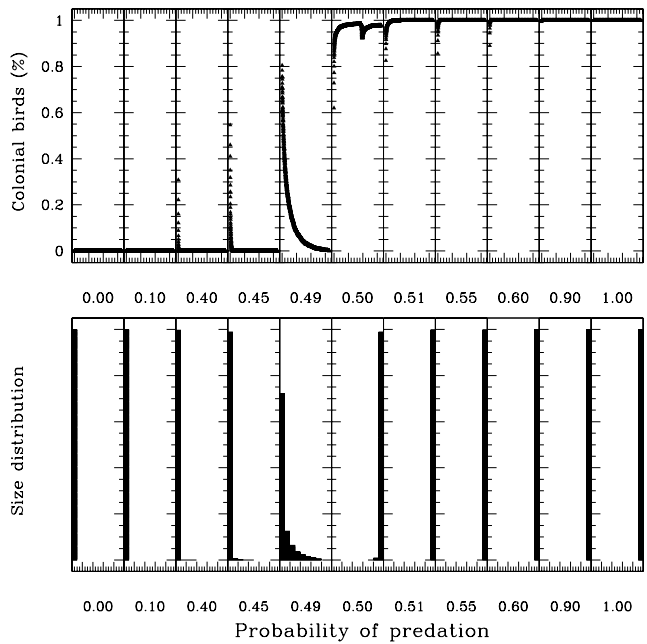}
\caption{Results for $m=2$, $s=5$, $N=10000$ and $v\rightarrow\infty$.
Top, temporal evolution of the model for predation
pressures for different predation pressures, from $t=1$ to $t=16384$.
Bottom, distribution
of colony sizes for the sam
e
temporal evolution. Note the change in the range of values
of $p$ studied with respect to those shown in fig.~\protect{\ref{f:mf1}}}
\label{f:mf2}
\end{figure}

It is interesting to consider the case of infinite lifespan,
although biologically unrealistic, shown in fig.~\ref{f:mf2}.
Then, the population reaches a stationary state where all birds
behave individually or colonially, with a sharp phase transition
at $p = 0.5$. This result can be obtained by estimating the
balance between costs and benefits of each type of behavior for
the whole population. Thus, the learning scheme described by this
model is guaranteed to lead to the optimal behavior if the
learning ability, or the memory, of the birds was infinity. Near
$p=0.5$ we find a very long lived transient, which tends to become
a power law decay, in line with the critical slowing down near a
second order phase transition~\cite{HH77}. In this sense, one
consider the stationary distribution at finite lifespans
(fig.~\ref{f:mf1}) as the equivalent of finite size effects near a
continuous phase transition.


\begin{table}
\begin{ruledtabular}
\begin{tabular}{lccc}
Choice & Success & $\Delta_i$& $\Delta_c$ \\
\hline
Individual &
\mbox{
\begin{tabular}{c}
Predated\\
Not Predated\\
\end{tabular}
}
&
\mbox{
\begin{tabular}{c}
0 \\
2 \\
\end{tabular}
}&
$1-2\times p/N$\\
\hline
Colonial &
\mbox{
\begin{tabular}{c}
Predated\\
Not Predated\\
\end{tabular}
}
&
$2\times (1-p)$
&
\mbox{
\begin{tabular}{c}
$1-2/n_{c}$ \\
1 \\
\end{tabular}
}
\\
\end{tabular}
\end{ruledtabular}
\caption{Collective scoring, where $N$ is the number of birds,
$p$ is the predation
probability and $n_{c}$ is the number of colonial birds.
$\Delta_i$ is the increase on the
score of an individual strategy and
$\Delta_c$ is the increase on the score of a colonial strategy.}
\label{t:mf}
\end{table}

\subsection{Individual scoring model\label{ss:indiv}}

It is equally or more consistent with existing field data to assume that
each bird makes all choices according to its own experiences.
This requires to modify
the scoring assigned to the strategies not followed by each
bird, defined in table~\ref{t:mf}.
The simplest choice is
to assume that, when unsuccesful (predated),
the bird assigns to the strategies leading to the
option not followed the score corresponding to the benefit
of that behavior in the absence of predation, as shown
in table~\ref{t:mf2}.
A bird who made a succesful choice (not predated),
updates only the scores of those strategies which lead
to that choice.
Finally, we assume that the predation pressure
is not the same for all colonial birds, as they form colonies
of different sizes. In order to take this into account, we
distribute the colonial birds into colonies which sizes
follow a power law distribution.
The number of colonies of
size $w$ is
proportional to $w^{-1}$.
This is the expected
behavior if the relative fluctuation of colony sizes is random.
The distribution is normalized to the number of colonial birds.

At each time step, which corresponds to one breeding season, there is a finite
probability $p$ that a nest will be predated.
Unless otherwise stated, the predation probability, $p$ is constant in time,
though the inclusion of variable probability does not change
the results qualitatively.

As in the previous case, individual strategies, when successful, obtain a
larger score than colonial strategies (2 vs. 1),
reflecting the innate (or genetic) preference of birds to individual breeding
in the absence of predation.

\begin{figure}
\includegraphics[width=\columnwidth]{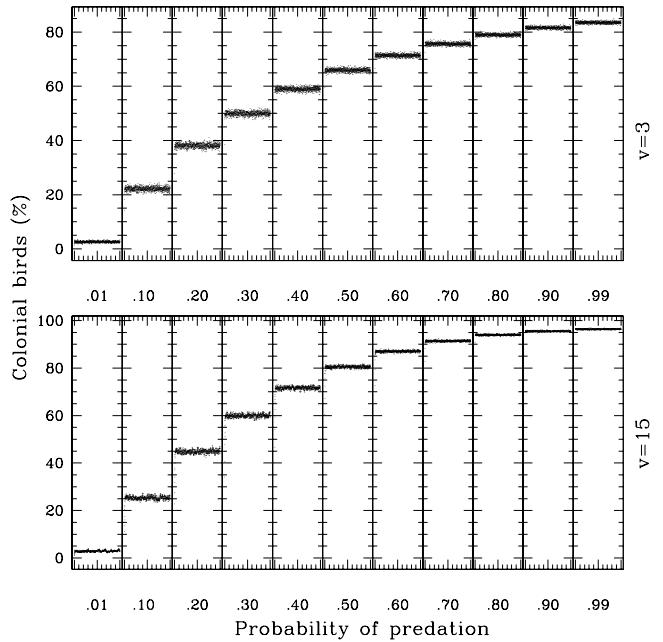}
\caption{Results for $m=1$, $s=4$, $N=10000$.
Temporal evolution of the individual scoring model from $t=1$ to $t=10000$,
in the cases of predation pressures from $p=0.01$ to $p=0.99$ ,
for $v=3$ (top) and $v=15$ (bottom).}
\label{f:ind3-15}
\end{figure}

We have studied the temporal evolution of the number (or frequency) of
colonial breeders, for different values of $m$, $s$ and $v$, performing
simulations for each combination of these three
parameters from $p=0.01$ to $p=0.99$.
In fig.~\ref{f:ind3-15} we can observe that the qualitative features
are equivalent for $v=3$ and $v=15$, however with longer lifespan,
birds can learn more, and therefore the adaptation process is clearer
for $v=15$. These results are in agreement with field studies~\cite{T96}.

\begin{table}
\begin{ruledtabular}
\begin{tabular}{lccc}
Choice & Success & $\Delta_i$& $\Delta_c$ \\
\hline
Individual &
\mbox{
\begin{tabular}{c}
Predated\\
Not Predated
\end{tabular}
}
&
\mbox{
\begin{tabular}{c}
0 \\
2 \\
\end{tabular}
}&
\mbox{
\begin{tabular}{c}
1 \\
0 \\
\end{tabular}
}
\\
\hline
Colonial &
\mbox{
\begin{tabular}{c}
Predated\\
Not Predated\\
\end{tabular}
}
&
\mbox{
\begin{tabular}{c}
2 \\
0 \\
\end{tabular}
}
&
\mbox{
\begin{tabular}{c}
0 \\
1 \\
\end{tabular}
}
\\
\end{tabular}
\end{ruledtabular}
\caption{Scoring in a Individual Model\label{t:ind}. $\Delta_i$ is the increase
on the score of a strategy that gives an individual outcome and $\Delta_c$ is the increase
on the score of a strategy that gives a colonial outcome.}
\label{t:mf2}
\end{table}

\subsection{Modelling colony distributions\label{ss:dist}}

Finally, we will consider explicitly the influence of the colony size
distribution, which is not taken as given. We start with a population
of individual birds, $n_b$, distributed among $n_s$ sites,
where $n_b \ll n_s$.
Birds have two possible strategies or behaviors, individual or colonial,
which have a score which reflects the reproductive success
that a bird would have if it would have followed it.
Note that we do not make use here of the set of the strategies
of the MG (such as in table~\ref{strategy}), but only these two stretegies,
as well as the information of the previous time step.
These birds are predated,
and use scoring rules similar to those
described in the previous subsection, and given in table~\ref{t:mf2}.
Birds which, at a given time step, choose to follow colonial
behavior, join an existing colony. At the beginning, as no colonies
still exist, birds which acquire this behavior are paired among them.
A given colonial bird has the same probability of joining any one of the
existing colonies~\cite{colonies}. Otherwise the bird (couple) nests
in one of the available empty sites.
When the score of a bird is taken to zero, the
(new) bird has an innate tendency towards individual behavior.
Finally, and
guided by field observations, we have considered the case where
birds make new choices every season, and the case where
birds which have not been predated repeat the previous choice, and only
consider their possible choices if they have been predated.

The model leads to different colony distributions. In general,
after some transients, large colonies appear and grow
indefinitely, leading to distributions skewed towards large sizes.
In this respect, the model differs from the two cases considered
previously. In the model discussed in~\ref{ss:mf}, the growth of
large colonies was arrested because birds were able to appraise
the collective performance of colonial birds. For the individual
model in~\ref{ss:indiv}, we assumed a fixed distribution
of possible colony sizes. By
combining an individual scoring procedure and not imposing
constraints on the distribution of colony sizes, we find that
the average size drifts towards large values, induced by the dilution of
predation presure for large colonies.

This tendency towards large colony sizes, for arbitrarily
small predation may help to explain the existence today of
species which form very large colonies, like penguins and
other sea birds. In many cases, however, there is an upper
limit to the maximum size that a colony can have, because
of the different disadvantageous effects of colonies,
such as parasitism, transmission of diseases,
lack of food on the vicinity of the colony
and the attraction of big predators to big colonies.
We have incorporated this possibility
by assuming that there are catastrophic events which
act on all members of a colony, and which probability
increases with the size of the colony. Similar effects are
obtained if the probability of predation includes the
possibility that there are events where a
whole colony is predated.

\begin{figure}
\includegraphics[width=\columnwidth]{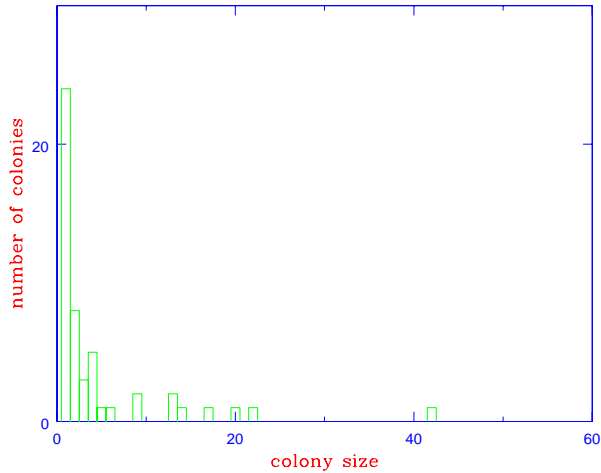}
\caption{Fit to the average colony distribution
for a population of approximately 300 couples
of lesser kestrel in Los Monegros (Spain)~\protect{\cite{T96}}.
The number of couples is 300, and the number of available
sites where colonies can be formed is 15000.
The predation pressure is $p = 0.08$, and there
is a finite probability of catastrophic events, $p_{cat}
= 0.01$, which limits the maximum colony size to $\sim 100$.}
\label{fit}
\end{figure}

The present model, including (few) catastrophic events which limit
the maximum colony size, allows us to fit
observed colony distributions~\cite{T96}. A fit to
results for colonies of lesser kestrels in Los
Monegros (Spain) is shown in figure~\ref{fit}.
The predation pressure is $p = 0.08$, the average
lifetime is $v = 5$, and the ratio between the success of
individual birds and colonial birds, in the absence of
predation, is $\Delta_i / \Delta_c = 2$. We also
assume a catastrophic predation $p_{cat} = 0.01$.
This implies that colonies cannot grow to sizes
much larger than 100.

When we introduce an upper cutoff the maximum colony size, the
numerical results are very suggestive of a rounded second order
phase transition, as in the cases discussed in~\ref{ss:mf}
and~\ref{ss:indiv}. We find a regime where most birds choose
individual strategies, for low predation pressure, and a regime
where most birds form colonies. The main difference with the
previous cases is that the critical predation pressure, $p_c$, at
which this transition takes place, is now lower. Our results
suggest that, in the present case, $p_c \approx 0.08$. This
reduction in the value of $p_c$ is consistent with the enhanced
tendency towards coloniality in this version of the model.
The evolution towards a stationary state is very slow, and,
for the parameters used,
at least 10$^3$ time steps are required.

\begin{figure}
\includegraphics[width=\columnwidth]{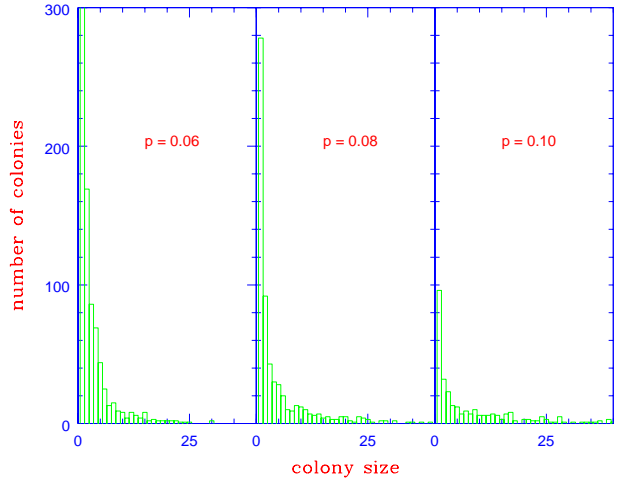}
\caption{Colony distributions obtained for a population of
3000 couples, and $p = 0.06$, left, $p = 0.08$, center,
and $p = 0.10$, right.}
\label{critical}
\end{figure}

It is interesting to note that the best value of $p$ which fits
the observed broad distribution of colony sizes is close to the
critical value which separates the two regimes described earlier.
This is best appreciated in fig.~\ref{critical}, where we have repeated the
calculations which lead to fig.~\ref{fit}, $p = 0.08$, and also
$p = 0.06$ and $p = 0.10$, for
a population of 3000 birds, and leaving all other parameters unchanged.
These results suggest that the assumption of a power law distribution
of colony sizes, made in~\ref{ss:mf} and~\ref{ss:indiv} is consistent.

\section{Conclusions.}
Our results suggest that colonial behavior can arise as a natural response
to predation pressure. Note that we made a number of
conservative assumptions, in order
to avoid any bias towards colonial behavior:
i) The birds have an innate tendency
towards preferring individual nests, ii) The only protection provided by the
colony is the dilution effect, iii) The distribution of colonies is such that
small colonies are more abundant, and, in some variations of the model,
%
iv) predation pressure fluctuates strongly
from year to year, allowing for the existence of periods of low predation.

The number of
colonial birds increases with increasing lifespan,
as birds accumulate experience for a longer period. This evidence is
in agreement with the observation that birds make use of their long term
breeding experiences\cite{Setal85,BG89,TH89,Oetal99}.

The models used here are inspired in the minority game
model, and use similar definitions of
allowed choices and strategies. On the other hand,
agents use their individual experiences in order to update
the scores of the different strategies, and the payoffs
are related to a random event, the chance of being
predated.

Our results suggest that simple mathematical models
of predation pressure on colonial birds can lead to a dynamical
phase transition, in which a majority of birds change from colonial
to individual breeding behavior. This transition is smoothed
due to the finite lifespan of the birds, which limits the
ability to learn new behaviors. Note, however, that the models
used in the present work cannot be expressed in terms of
the minimization of a benefit function.

Finally, it is intriguing that field observations~\cite{T96} can be fitted
by the model in~\ref{ss:dist} by tuning the parameters to be close to the
critical point discussed above, suggesting some kind of
self critical organization~\cite{J00}.
\section{Acknowledgements.}
We are thankful to F. Hiraldo, J. A. Don\'azar, M. G. Forero,
J. Cartwright, J. M. Garc{\'\i}a Ruiz and F. Ot\'alora
for helpful discussions. We acknowledge financial support
from grants PB96-0875 (MCyT, Spain), 07N/0045/98 (C. Madrid),
and Caja de Ahorros de Granada ``La General''.


\begin{thebibliography}{99}
\bibitem{colonies}
We have also used more realistic models, in which birds are able
to judge the well being
of a given colony, by weighting each colony
by a factor proportional to the fraction of birds which were
not predated. These changes in the model do not alter qualitatively
the results.
\end{thebibliography}
\end{document}